 \documentclass[final,3p,times,twocolumn]{elsarticle}

\usepackage{amssymb}

\biboptions{sort&compress}

\journal{NIM B}

\begin{document}

\begin{frontmatter}



\title{Activation cross-sections of deuteron induced nuclear reactions on neodymium up to 50 MeV}


\author[1]{F. T\'ark\'anyi}
\author[1]{S. Tak\'acs}
\author[1]{F. Ditr\'oi\corref{*}}
\author[2]{A. Hermanne}
\author[3]{H. Yamazaki} 
\author[3]{M. Baba}
\author[3]{A. Mohammadi} 
\author[4]{A.V. Ignatyuk}
\cortext[*]{Corresponding author: ditroi@atomki.hu}

\address[1]{Institute for Nuclear Research, Hungarian Academy of Sciences (ATOMKI),  Debrecen, Hungary}
\address[2]{Cyclotron Laboratory, Vrije Universiteit Brussel (VUB), Brussels, Belgium}
\address[3]{Cyclotron Radioisotope Center (CYRIC), Tohoku University, Sendai, Japan}
\address[4]{Institute of Physics and Power Engineering (IPPE), Obninsk, Russia}

\begin{abstract}
In the frame of a systematic study of activation cross sections of deuteron induced nuclear reactions on rare earths, the reactions on neodymium for production of therapeutic radionuclides were measured for the first time. The excitation functions of the $^{nat}$Nd(d,x) $^{151,150,149,148m,148g,146,144,143}$Pm, $^{149,147,139m}$Nd, $^{142}$Pr and $^{139g}$Ce nuclear reactions were assessed by using the stacked foil activation technique and high resolution $\gamma$-spectrometry. The experimental excitation functions were compared to the theoretical predictions calculated with the modified model codes ALICE-IPPE-D and EMPIRE-II-D and with the data in the TENDL-2012 library based on latest version of the TALYS code. The application of the data in the field of medical isotope production and nuclear reaction theory is discussed.
\end{abstract}


\end{frontmatter}


\section{Introduction}
\label{1}
Activation cross-sections data of deuteron induced nuclear reactions on neodymium are important for development of nuclear reaction theory and for different practical applications.  
This study was performed in the frame of the following ongoing research goals:
\begin{itemize}
\item To check the predictive power and benchmarking of the different model codes of the nuclear reaction theory. To contribute to the improvement of models for description of deuteron induced reactions and to selection of more appropriate input parameters.
\item To investigate the alternative production possibilities of standard and emerging therapeutic and diagnostic lanthanide radionuclides via charged particle induced reactions. In a search for new nuclides suitable for therapeutic purposes \cite{1,2,3,4,5,6,7,8} the radionuclide $^{149}$Pm (T$_{1/2}$ = 53.1 h), $^{141}$Nd (T$_{1/2}$) and $^{140}$Nd (T$_{1/2}$ = 3.37 d),  were found to offer some unique properties suitable for therapy  and  the daughter nuclide $^{140}$Pr (T$_{1/2}$ = 3.4 min) brings the additional advantage of in-vivo localization via positron emission tomography (PET).
\item To contribute to the extension of the database by a systematic study of activation cross-sections of deuteron induced nuclear reactions for biological and industrial applications (accelerator technology, thin layer activation, activation in space technology, etc.).
\end{itemize}
Here we present our results on activation cross-sections on deuteron induced nuclear reactions on neodymium. No earlier experimental data on $^{nat}$Nd were found in the literature. Only one earlier set of experimental cross-section data was found on highly enriched $^{148}$Nd for investigation the isomeric ratios of the (d,2n) reactions \cite{9}.  
Thick target yield data for production of $^{143,144,148}$Pm at 22 MeV were reported by Dmitriev et al.  at  22 MeV \cite{10}.

\section{Experiment and data evaluation}
\label{2}
For measurements, the well-known activation method, stacked foil irradiation technique and high resolution $\gamma$-spectrometry was used. Nd metal foils and NdO pellet targets, interleaved with Al foils for monitoring of beam characteristics, were stacked and irradiated at CYRIC (Sendai) and UCL (LLN) cyclotrons. Complete excitation function was measured for the monitor reactions to control the beam intensity and the energy. The main experimental parameters, methods of data evaluation are collected in Table 1 and Table 2.  The comparison of the re-measured monitor reactions and the recommended data are shown in Fig. 1. The decay characteristics of the investigated reaction products and the possibly contributing reactions in the energy region studied are summarized in Table 3.

\begin{figure}[h]
\includegraphics[scale=0.3]{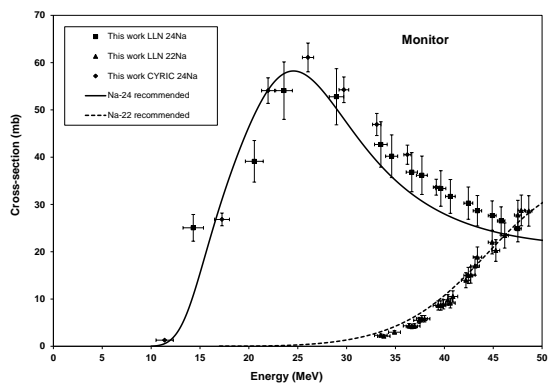}
\caption{Comparison of the measured and the recommended cross-sections or the used monitor $^{27}$Al(d,x)$^{22,24}$Na reactions }
\end{figure}

\begin{table*}[t]
\tiny
\caption{Main experimental parameters}
\centering
\begin{center}
\begin{tabular}{|p{1.3in}|p{1.3in}|p{1.6in}|} \hline 
Incident particle & Deuteron  & Deuteron  \\ \hline 
Method  & Stacked foil & Stacked foil \\ \hline 
Target composition & ${}^{nat}$Nd (100 $\mu$m)-target\newline ${}^{nat}$Yb (22.88 $\mu$m)-target\newline  Al \textbf{(}49.06 $\mu$m)-monitor\newline (repeated 11 times) \newline  Interleaved with Al (156.6 $\mu$m, 103.43 $\mu$m, 49.06 $\mu$m)-energy degraders\newline \newline  & 74.7-93.7 $\mu$m Nd${}_{2}$O${}_{3}$ target  \newline pressed into 360 $\mu$m thick Al backing covered  by 10 $\mu$m Al foil \newline  \\ \hline 
Number of Nd  target foils & 11 & 8 \\ \hline 
Accelerator & Cyclone 110 cyclotron of the Université Catholique in Louvain la Neuve (LLN) Belgium & AVF-930 cyclotron of the Cyclotron and Radioisotope Center (CYRIC)\underbar{ }of\textbf{ }Tohoku University in Sendai,\newline Japan \\ \hline 
Primary energy & 50 MeV & 40 MeV \\ \hline 
Covered energy range & 49.4-21.5 MeV & 39.4-9.7 MeV \\ \hline 
Irradiation time & 60  min & 20 min \\ \hline 
Beam current & 92 nA & 45 nA \\ \hline 
Monitor reaction, [recommended values]  & ${}^{27}$Al(d,x)${}^{24}$Na  reaction [11] (re-measured over the whole energy range) & ${}^{27}$Al(d,x)${}^{24}$Na  reaction\newline (re-measured over the whole energy range) \\ \hline 
Monitor target and thickness & ${}^{nat}$Al, 49.06 mm & ${}^{nat}$Al, 370 mm target holders \\ \hline 
detector & HPGe & HPGe \\ \hline 
Chemical separation & no & no \\ \hline 
g-spectra measurements & 4 series & 3 series \\ \hline 
Typical cooling times\newline (and target-detector distances) & 3 h (35 cm)\newline 34 h (10 cm)\newline 47 h (20 cm)\newline 272 (5cm)\newline  & 19 h (10 cm)\newline 90 h (10 cm)\newline  600 h (10 cm) \\ \hline 
\end{tabular}
\end{center}

\end{table*}

\begin{table*}[t]
\tiny
\caption{Main parameters and methods of the data evaluation (with references)}
\centering
\begin{center}
\begin{tabular}{|p{1.4in}|p{2.1in}|p{0.6in}|} \hline 
Gamma spectra evaluation & Genie 2000, Forgamma & \cite{12, 13} \\ \hline 
Determination of beam intensity & Faraday cup (preliminary)\newline Fitted monitor reaction (final) &  \cite{14} \\ \hline 
Decay data (see Table 2) & NUDAT 2.6  & \cite{15} \\ \hline 
Reaction Q-values (see Table 3) & Q-value calculator & \cite{16} \\ \hline 
Determination of beam energy & Andersen (preliminary)\newline Fitted monitor reaction (final) & \cite{17}\newline \cite{15} \\ \hline 
Uncertainty of energy & Cumulative effects of possible uncertainties\newline (primary energy, target thickness, energy straggling,  correction to monitor reaction) &  \\ \hline 
Cross-sections & Isotopic and elemental cross-sections &  \\ \hline 
Uncertainty of cross-sections & sum in quadrature of all individual contributions\newline (beam current  (7\%), \newline beam-loss corrections  (max.  of  1.5\%),  \newline target thickness  (1\%),  \newline detector efficiency (5\%),  \newline photo peak area  determination  \newline counting statistics (1-20 \%)   & \cite{18} \\ \hline 
Yield & Physical yield & \cite{19} \\ \hline 
\end{tabular}
\end{center}

\end{table*}

\section{Comparison with the results of the model codes}
\label{3}
The cross-sections of the investigated reactions were calculated using the modified model codes ALICE-IPPE [20] and EMPIRE-II \cite{21}. In the used modified ALICE-IPPE-D and EMPIRE-D code versions the direct (d,p) channel is increased strongly \cite{22,23}. The experimental data are also compared with the cross-section data reported in the TALYS based \cite{24} TENDL-2012 Data Libraries \cite{25}.

\section{Results}
\label{4}

\subsection{Cross-sections}
\label{4.1}

The cross-sections for all the reactions investigated are shown in Figs. 2–15 and the numerical values are collected in Table 4-5. The results of the two irradiations on different targets are given separately allowing to judge on the agreement in the overlapping energy range. We should mention, as is reflected in the figures that the experimental results of this investigations have larger uncertainties and the data are more scattered, than in most of our previous investigations.  Taking into accounts that the scattering shows a systematic behavior between the two experiments, it could be connected to the target thicknesses. In the LLN experiment a large metal foil was used and the average thickness was derived from measurements of the surface of the whole foil and its weight.  The final targets were obtained by cutting pieces of the required dimensions from the large foil.  Due to the well-known oxidation problems the thickness of each individual target piece were not re-measured but supposed to be equal to the average. In the Sendai experiment the Nd2O3 targets were made by pressing neodymium oxide in a precisely machined Al cup. The average thickness of each individual targets was derived on the basis the measured oxide weight (total weight minus weight of Al-cup) and known surface, but the uniformity of the targets was only roughly checked. For both experiments the beam diameter was significantly lower than the diameter of the targets and hence possible local changes in thickness can result in large scattering of cross-section data.

\begin{table*}[t]
\tiny
\caption{Decay characteristic of the investigated reaction products and the contributing reactions}
\centering
\begin{center}
\begin{tabular}{|p{0.8in}|p{0.6in}|p{0.6in}|p{0.5in}|p{0.8in}|p{0.8in}|} \hline 
Nuclide\newline Decay path & Half-life & E${}_{\gamma}$(keV) & I${}_{\gamma}$(\%) & Contributing reaction & Q-value\newline (keV) \\ \hline 
\textbf{${}^{151}$Pm\newline }~$\beta $${}^{-}$: 100 \%~\textbf{} & 28.40 h & 240.09\newline 340.08\newline 445.68 & 3.8\newline 22.5 \newline 4.0  & ${}^{150}$Nd(d,n) & ~4770.71 \\ \hline 
\textbf{${}^{150}$Pm\newline }$\beta $${}^{-}$: 100\%~\textbf{} & 2.68 h & 333.92\newline ~406.51\newline 831.85\newline 876.41\newline 1165.77\newline 1324.51~ & ~68\newline ~5.6\newline ~11.9\newline 7.3\newline 15.8\newline 17.5 & ${}^{150}$Nd(d,2n) & -3089.5~ \\ \hline 
\textbf{${}^{149}$Pm\newline }b${}^{-}$: 100\%~~\textbf{} & 53.08~h & 285.95~ & 3.1~ & ${}^{1}$${}^{48}$Nd(d,n)\newline ${}^{150}$Nd(d,3n) & 3720.27\newline ~-8693.63~ \\ \hline 
\textbf{${}^{148m}$Pm\newline }IT: 4.2 \%\newline ~$\beta $${}^{-}$: 95.8 \%\textbf{} & 41.29 d & 432.745\newline 550.284~\newline 629.987\newline 725.673\newline 915.331~~\newline 1013.808~ & 5.33~\newline 94.5~\newline 89~\newline 32.7\newline 17.10~\newline 20.20~ & ${}^{1}$${}^{48}$Nd(d,2n)\newline ${}^{150}$Nd(d,4n) & ~-3549.66\newline ~-15963.56 \\ \hline 
\textbf{${}^{148g}$Pm\newline }$\beta $${}^{-}$: 100 \%\textbf{~} & 5.368 d & ~550.27\newline ~914.85\newline 1465.12 & 22.0\newline 11.5\newline 22.2 & ${}^{1}$${}^{48}$Nd(d,2n)\newline ${}^{150}$Nd(d,4n) & ~-3549.66\newline ~-15963.56 \\ \hline 
\textbf{${}^{146}$Pm\newline }$\varepsilon $: 66.0  \%~\textbf{} & ~5.53 a & 453.88\newline ~735.93 & 65.0\newline 22.5 & ${}^{1}$${}^{45}$Nd(d,n)\newline ${}^{1}$${}^{46}$Nd(d,2n)\newline ${}^{1}$${}^{48}$Nd(d,4n)\newline ${}^{150}$Nd(d,6n) & ~3086.77\newline -4478.47~\newline -17103.16\newline ~-29517.07 \\ \hline 
\textbf{${}^{144}$Pm\newline }$\varepsilon $: 100 \%\textbf{~} & 363 d & ~~476.78\newline 618.01\newline 696.49 & ~43.8\newline ~98\newline ~99.49 & ${}^{1}$${}^{43}$Nd(d,n)\newline ${}^{1}$${}^{44}$Nd(d,2n)\newline ${}^{1}$${}^{45}$Nd(d,3n)\newline ${}^{1}$${}^{46}$Nd(d,4n)\newline ${}^{1}$${}^{48}$Nd(d,6n)\newline ${}^{150}$Nd(d,8n) & 2478.19\newline ~-5338.84\newline ~-11094.15\newline ~-18659.38\newline -31284.08~\newline ~-43697.98 \\ \hline 
\textbf{${}^{143}$Pm\newline }$\varepsilon $: 100 \%\textbf{~} & ~265 d & ~741.98 & ~38.5 & ${}^{1}$${}^{42}$Nd(d,n)\newline ${}^{1}$${}^{43}$Nd(d,2n)\newline ${}^{1}$${}^{44}$Nd(d,3n)\newline ${}^{1}$${}^{45}$Nd(d,4n)\newline ${}^{1}$${}^{46}$Nd(d,5n)\newline ${}^{1}$${}^{48}$Nd(d,7n)\newline ${}^{150}$Nd(d,9n) & ~~2074.99\newline ~-4048.59\newline ~-11865.62\newline -17620.92\newline -36201.1\newline ~-37810.85\newline ~-50224.76 \\ \hline 
\textbf{${}^{149}$Nd\newline }$\beta $${}^{-}$: 100 \%\textbf{} & 1.728 h & 114.314\newline 211.309\newline ~270.166\newline 326.554\newline ~423.553\newline 540.509\newline 654.831 & 19.2\newline 25.9\newline 10.7\newline ~4.56\newline 7.4\newline 6.6\newline 8.0  & ${}^{1}$${}^{48}$Nd(d,p)\newline ${}^{150}$Nd(d,p2n)\newline ${}^{149}$Pr decay & 2814.224\newline -9599.69\newline ~-12153.2~ \\ \hline 
\textbf{${}^{147}$Nd\newline }$\beta $${}^{-}$: 100 \%\textbf{} & 10.98 d & 91.105\newline 319.411\newline 531.016 & 28.1\newline 2.127\newline ~13.37 & ${}^{1}$${}^{46}$Nd(d,p)\newline ${}^{1}$${}^{48}$Nd(d,p2n)\newline ${}^{150}$Nd(d,p4n)\newline ${}^{14}$${}^{7}$Pr decay & ~3067.6343\newline -9557.07\newline ~-21970.98\newline ~4404.4~ \\ \hline 
\textbf{${}^{141}$Nd\newline }~$\varepsilon $: 100 \%~\textbf{} & 2.49 h & 145.45\newline ~1126.91\newline 1147.30\newline 1292.64 & 0.24\newline 0.80\newline 0.307\newline ~0.46 & ${}^{1}$${}^{42}$Nd(d,p2n)\newline ${}^{1}$${}^{43}$Nd(d,p3n)\newline ${}^{1}$${}^{44}$Nd(d,p4n)\newline ${}^{1}$${}^{45}$Nd(d,p5n)\newline ${}^{1}$${}^{46}$Nd(d,p6n)\newline ${}^{1}$${}^{48}$Nd(d,p8n)\newline ${}^{150}$Nd(d,p10n)\newline ${}^{14}$${}^{1}$Pm decay & -12052.37~\newline -18175.95\newline -25992.98\newline ~-31748.28\newline -39313.52\newline -51938.22~\newline -65222.22\newline -16505.2\newline  \\ \hline 
\textbf{${}^{139m}$Nd\newline }$\varepsilon $: 88.2\%\newline IT: 11.8\%\textbf{\newline } & 5.50 h & 113.87\newline 708.1\newline 738.2\newline 827.8\newline 982.2 & 40\newline 26\newline ~35\newline 10.3\newline 26 & ${}^{1}$${}^{42}$Nd(d,p4n)\newline ${}^{1}$${}^{43}$Nd(d,p5n)\newline ${}^{1}$${}^{44}$Nd(d,p6n)\newline ${}^{1}$${}^{45}$Nd(d,p7n)\newline ${}^{1}$${}^{46}$Nd(d,p8n)\newline ${}^{1}$${}^{39}$Pm decay & ~-30373.8~\newline -36497.3\newline -44314.4\newline ~-50069.7\newline ~-57634.9\newline -35670.2 \\ \hline 
\textbf{${}^{142}$P}r\newline ~$\varepsilon $: 0.0164\%\newline $\beta $${}^{-}$: 99.9836\% & 19.12 h & 1575.6 & 3.7 & ${}^{1}$${}^{42}$Nd(d,2p)\newline ${}^{1}$${}^{43}$Nd(d,2pn)\newline ${}^{1}$${}^{44}$Nd(d,2p2n)\newline ${}^{1}$${}^{45}$Nd(d,2p3n)\newline ${}^{1}$${}^{46}$Nd(d,2p4n)\newline ${}^{1}$${}^{48}$Nd(d,2p6n)\newline ${}^{150}$Nd(d,2p8n) & ~~-3603.85\newline -9727.43\newline ~-17544.46\newline ~-23299.77\newline ~-30865.0\newline ~~-43489.7\newline ~-55903.59 \\ \hline 
\textbf{${}^{139}$Ce\newline }e: 100\% & 137.641 d & 165.8575 & 80 & ${}^{1}$${}^{42}$Nd(d,3p2n)\newline ${}^{1}$${}^{43}$Nd(d,3p3n)\newline ${}^{1}$${}^{44}$Nd(d,3p4n)\newline ${}^{1}$${}^{45}$Nd(d,3p5n)\newline ${}^{1}$${}^{46}$Nd(d,3p6n)\newline ${}^{1}$${}^{48}$Nd(d,3p8n)\newline ${}^{139}$Pr decay & -23873.51\newline -29997.09\newline -37814.12~\newline -43569.42\newline ~-51134.66\newline ~-63759.34\newline 1510.73 \\ \hline 
\end{tabular}

\end{center}
\begin{flushleft}
\footnotesize{When complex particles are emitted instead of individual protons and neutrons the Q-values have to be decreased by the respective binding energies of the compound particles: np-d, +2.2 MeV; 2np-t, +8.48 MeV} \\
${}^{nat}$Nd isotopic abundance:  ${}^{14}$${}^{2}$Nd (27.13 \%), ${}^{14}$${}^{3}$Nd (12.18 \%), ${}^{14}$${}^{4}$Nd (23.80 \%),${}^{14}$${}^{5}$Nd (5.30 \%), ${}^{14}$${}^{6}$Nd (17.19 \%), ${}^{14}$${}^{8}$Nd (5.76 \%), ${}^{1}$${}^{50}$Nd (5.64 \%)
\end{flushleft}

\end{table*}

\subsubsection{ Production cross-sections of  $^{151}$Pm (cum)}
\label{4.1.1}
The $^{151}$Pm (T$_{1/2}$=28.40 h) is produced directly via $^{150}$Nd(d,n) reaction and from the decay of  short-lived $^{151}$Nd (T$_{1/2}$ = 12.44 min) obtained in a $^{150}$Nd(d,p) reaction. Due to the experimental circumstances we could only measure the cumulative production after the complete decay of the short-lived parent. As it is shown in Fig. 2 the experimental data are in acceptable agreement with the results of ALICE-D and EMPIRE-D, but in case of TENDL-2012 the underestimation is very significant.

\begin{figure}
\includegraphics[scale=0.3]{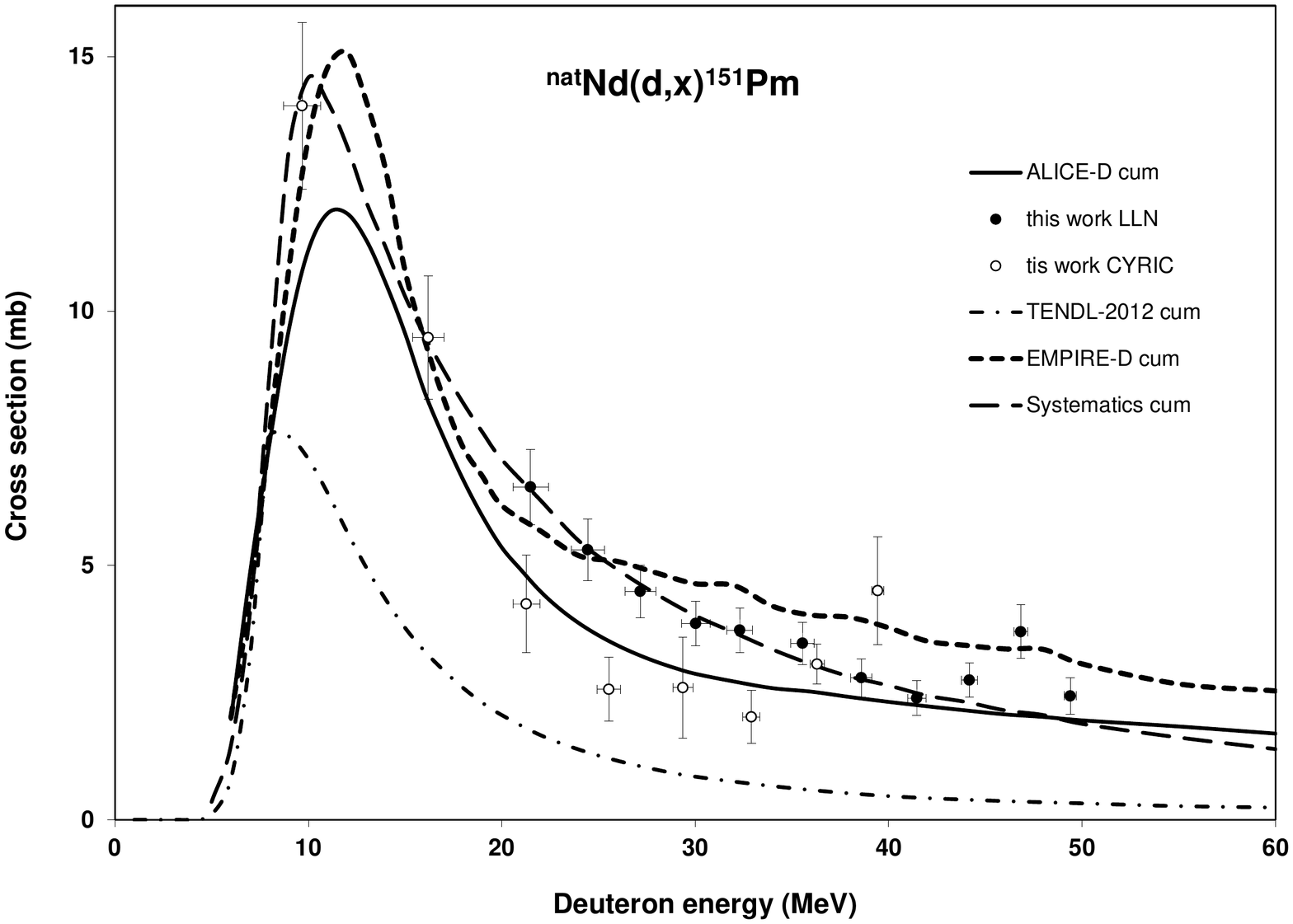}
\caption{Experimental cross-sections for the $^{nat}Nd$(d,x)$^{151}$Pm reaction and comparison with the theoretical code calculations}
\end{figure}

\subsubsection{Production cross-sections of  $^{150}$Pm}
\label{4.1.2}
 For $^{150}$Pm (T$_{1/2}$ = 2.68 h)  we have experimental data only from the 50 MeV irradiation due to the long cooling time (19 h) of the first series of measurement in the 40 MeV experiment. The $^{150}$Pm can only be produced via the $^{150}$Nd(d,2n) reaction. The measured data and the theoretical model calculations are shown in Fig. 3. The agreement between the experimental and theoretical results is acceptable.

\begin{figure}
\includegraphics[scale=0.31]{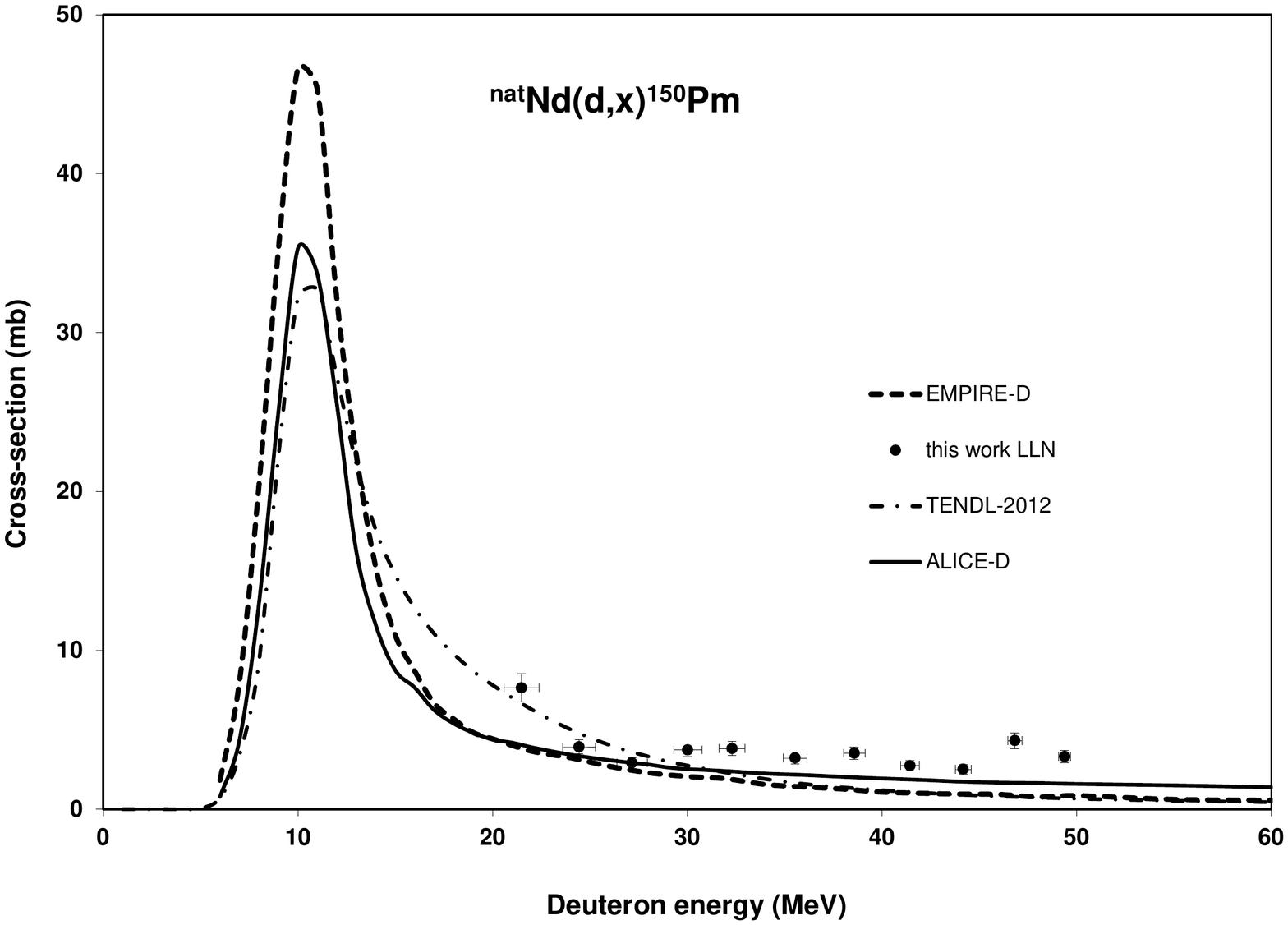}
\caption{Experimental cross-sections for the $^{nat}Nd$(d,x)$^{150}$Pm reaction and comparison with the theoretical code calculations}
\end{figure}

\subsubsection{Production cross-sections of  $^{149}$Pm}
\label{4.1.3}
The experimental and theoretical excitation functions of $^{149}$Pm  (T$_{1/2}$ = 53.08 h) are shown in Fig. 4.  Two reactions are contributing on stable Nd isotopes with approximately the same natural abundance ($^{148}$Nd(d,n) and $^{150}$Nd(d,3n)). The maximum around 20 MeV (seen in the experiment and confirmed by the theoretical predictions) indicates that the cross-sections for the (d,3n) reaction are significantly higher than those for (d,n).

\begin{figure}
\includegraphics[scale=0.31]{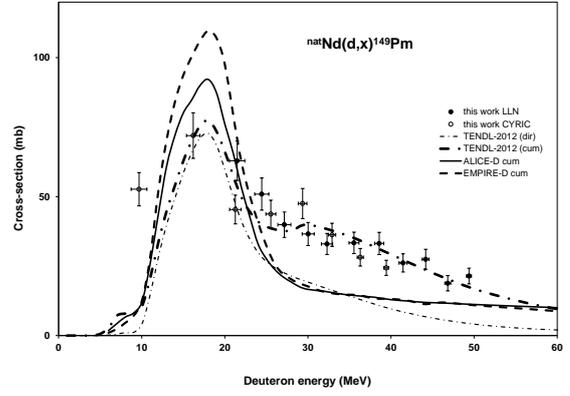}
\caption{Experimental cross-sections for the $^{nat}$Nd(d,x)$^{149}$Pm reaction and comparison with the theoretical code calculations}
\end{figure}

\subsubsection{Production cross-sections of  $^{148m}$Pm}
\label{4.1.4}

The measured excitation functions came in all cases from a combination of (d,xn) reactions on the $^{185}$Re and $^{187}$Re. The level schemes for $^{182}$Re and $^{184}$Re are estimated with large uncertainties for some important  $\gamma$-transitions. So, an accuracy of calculations for the corresponding isomer yields cannot be very high.

The $^{148}$Pm has two long-lived isomeric states. The $^{148m}$Pm (T$_{1/2}$ = 41.29 d) higher laying state shows only for a small part isomeric decay (IT: 4.2 \%) and emits strong, independent $\gamma$-lines allowing separate identification.  The agreement of our 2 data points below 18 MeV with the data of \cite{9} is reasonable. The theoretical data overestimate the experimental values (Fig. 5). The maxima of the two contributing reactions ((d,2n) and (d,4n)) are well pronounced in the figure, especially for the model calculations.

\begin{figure}
\includegraphics[scale=0.31]{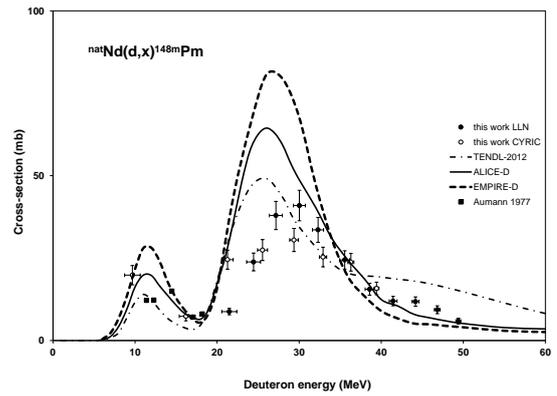}
\caption{Experimental cross-sections for the $^{nat}$Nd(d,x)$^{148m}$Pm reaction and comparison with the theoretical code calculations}
\end{figure}

\subsubsection{Production cross-sections of  $^{148g}$Pm}
\label{4.1.5}
The cross-sections for the direct production of the $^{148g}$Pm (T$_{1/2}$ = 5.368 d) (after correction for the small contribution of the $^{148m}$Pm decay) are shown in Fig. 6 in comparison with the earlier experimental data and the theoretical predictions. The overestimation by the theory for production of the ground state is significant. Our experimental results below 18 MeV for production of the ground and metastable state are not detailed enough to allow any calculation of isomeric ratio and comparison with \cite{9}.

\begin{figure}
\includegraphics[scale=0.31]{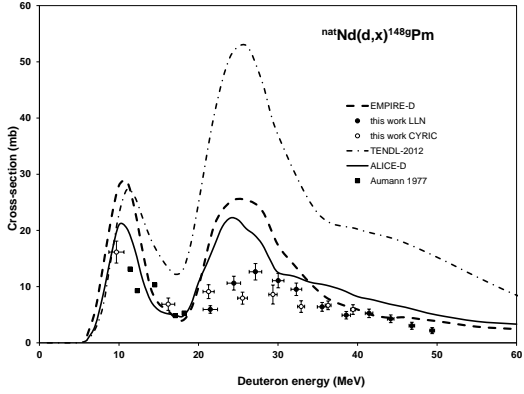}
\caption{Experimental cross-sections for the $^{nat}$Nd(d,x)$^{148g}$Pm reaction and comparison with the theoretical code calculations}
\end{figure}

\subsubsection{Production cross sections of  $^{147}$Pm}
\label{4.1.6}
Due to the long half-life (T$_{1/2}$ = 2.62 a) and the very low abundance of its unique -line this industrially important radionuclide could not be detected, neither as directly produced nor as decay product of its parent $^{147}$Nd.

\subsubsection{Production cross-sections of  $^{146}$Pm}
\label{4.1.7}
The comparison of the experimental and the theoretical data of the $^{146}$Pm (T$_{1/2}$ = 5.53 a) shows significant overestimation of the experiment by the used models that however clearly show the contributions of different stable target isotopes (Fig. 7).

\begin{figure}
\includegraphics[scale=0.31]{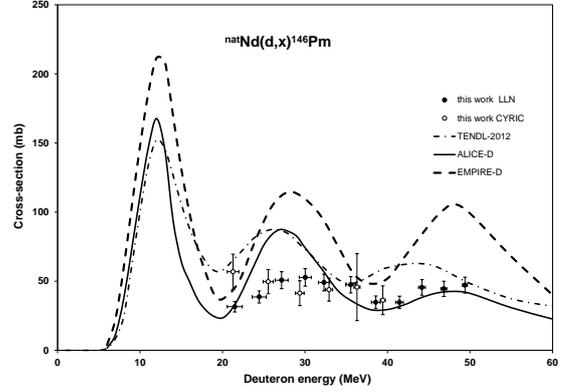}
\caption{Experimental cross-sections for the $^{nat}$Nd(d,x)$^{146}$Pm reaction and comparison with the theoretical code calculations}
\end{figure}

\subsubsection{Production cross-sections of  $^{144}$Pm}
\label{4.1.8}
The $^{144}$Pm (T$_{1/2}$ = 363 d) is produced by (d,xn) reactions on 143-$^{150}$Nd  isotopes of $^{nat}$Nd. The shape of the excitation function predicted in TENDL-2012 differs significantly from the predictions of the ALICE-D and EMPIRE-D and from the experiment, especially at the high energy range (Fig. 8).

\begin{figure}
\includegraphics[scale=0.31]{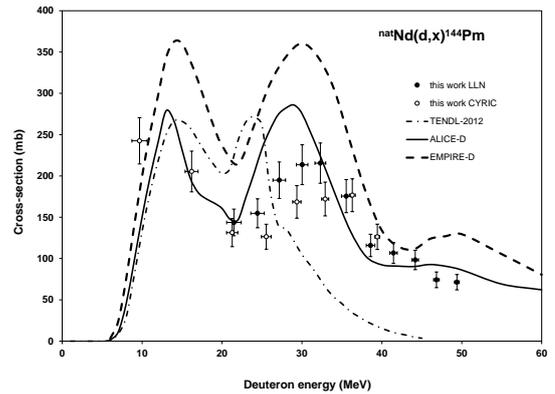}
\caption{Experimental cross-sections for the $^{nat}$Nd(d,x)$^{144}$Pm reaction and comparison with the theoretical code calculations}
\end{figure}

\subsubsection{Production cross-sections of  $^{143}$Pm}
\label{4.1.9}
The (d,xn) reactions on all stable isotopes of $^{nat}$Nd  participate in the production of the $^{143}$Pm ( T$_{1/2}$ = 265 d). The comparison of the experiment and the theory shows a similar picture as for previous cases: the experimental data are overestimated by the theory and are not distinguishing well the different contributing reactions (Fig. 9).

\begin{figure}
\includegraphics[scale=0.31]{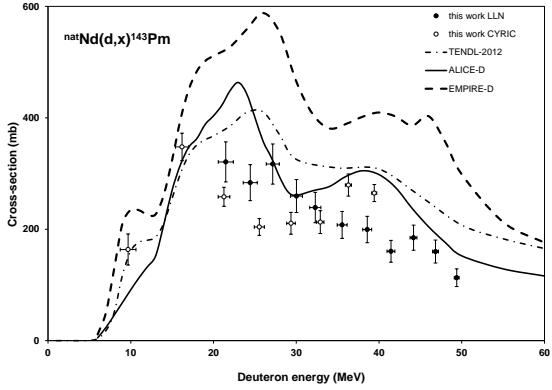}
\caption{Experimental cross-sections for the $^{nat}$Nd(d,x)$^{143}$Pm reaction and comparison with the theoretical code calculations}
\end{figure}

\subsubsection{Production cross-sections of  $^{149}$Nd(cum)}
\label{4.1.10}
Due to the long cooling time for the 40 MeV experiment (around 19 h) we could not measure the cross-sections for production of $^{149}$Nd (T$_{1/2}$ = 1.728 h) in this experiment.  The measured cross-sections for the 50 MeV experiment (Fig. 10) represent cumulative production of $^{149}$Nd directly by (d,pxn) and from the complete $\beta^-$-decay of the short-lived $^{149}$Pr (T$_{1/2}$ = 2.26 min). The two maxima seen on the theoretical excitation functions correspond to the $^{148}$Nd(d,p) and $^{150}$Nd(d,p2n) reactions. The contribution from the decay of $^{149}$Pr is negligible, according to the TENDL-2012 results and is only visible at energies above 50 MeV. There are large disagreements in the theoretical predictions. The predictions of the (d,p) part are more reliable in case of ALICE-D and EMPIRE-D, but the (d,p2n) energy range is strongly underestimated by these codes. In the measured energy region the TENDL-2012 results are more close to the experiment.

\begin{figure}
\includegraphics[scale=0.31]{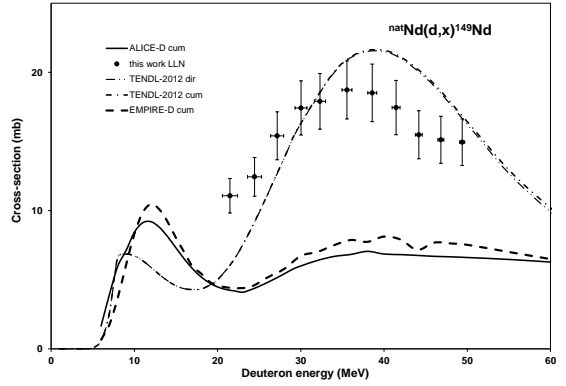}
\caption{Experimental cross-sections for the $^{nat}$Nd(d,x)$^{149}$Nd reaction and comparison with the theoretical code calculations}
\end{figure}

\subsubsection{Production cross-sections of  $^{147}$Nd(cum)}
\label{4.1.11}
The production cross-sections of $^{147}$Nd (T$_{1/2}$ = 10.98 d) were deduced from spectra taken after complete decay of short-lived parent $^{147}$Pr (T$_{1/2}$ = 13.4 min) and hence represent cumulative production (direct + decay from parent). The first maximum (Fig. 11) represents the (d,p) contribution on $^{146}$Nd, where the description of the ALICE-D and EMPIRE-D is better. It is difficult to compare the experimental data with the theory in the higher energy range.

\begin{figure}
\includegraphics[scale=0.31]{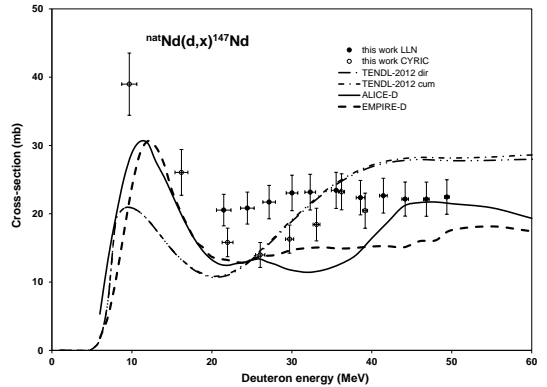}
\caption{Experimental cross-sections for the $^{nat}$Nd(d,x)$^{147}$Nd reaction and comparison with the theoretical code calculations}
\end{figure}

\subsubsection{Production cross-sections of  $^{141}$Nd(cum)}
\label{4.1.12}
The short-lived isomeric state (T$_{1/2}$ = 62.0 s) of $^{141}$Nd is decaying with IT (100 \%) to the ground state (T$_{1/2}$ = 2.49 h). The ground state of $^{141}$Nd is also fed by EC + $\beta^+$-decay of $^{141}$Pm (20.90 min) and is produced directly by (d,pxn) reactions on all stable Nd isotopes. The measured cross-sections for the production of the ground state are cumulative, including all above mentioned contributions (Fig. 12).  According to the Fig. 12 there are very large disagreements between the predictions of the different theoretical codes and the experimental data.

\begin{figure}
\includegraphics[scale=0.31]{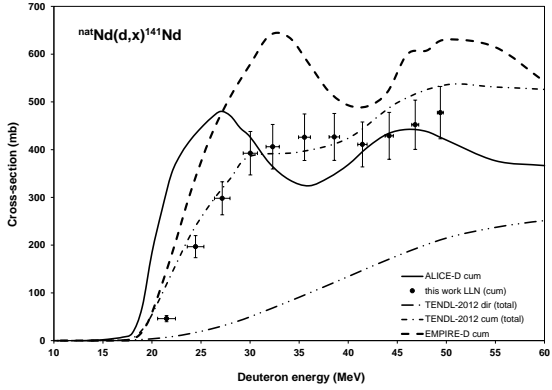}
\caption{Experimental cross-sections for the $^{nat}$Nd(d,x)$^{141}$Nd reaction and comparison with the theoretical code calculations}
\end{figure}

\subsubsection{Production cross-sections of  $^{139m}$Nd}
\label{4.1.13}
The $^{139}$Nd has two long-lived isomeric states, the ($I^{\pi} = 3/2^+$, T$_{1/2}$  =  29.7 min) ground state and the longer-lived high spin metastable state ($I^{\pi} =  11/2^-$,  T$_{1/2}$  =  5.50 h). We obtained cross-sections for production of the high spin state that is produced only directly via (d,pxn) reactions. The $^{139}$Pm ($I^{\pi} =5/2^+$, T$_{1/2}$  =  4.15 min) decays only to the ground state of 139Nd. Our experimental data (Fig. 13) are closer to the TALYS predictions in the TENDL-2012 than to the ALICE-D and EMPIRE-D results.

\begin{figure}
\includegraphics[scale=0.31]{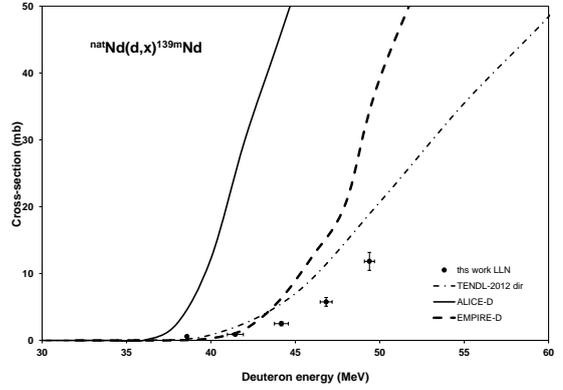}
\caption{Experimental cross-sections for the $^{nat}$Nd(d,x)$^{139m}$Nd reaction and comparison with the theoretical code calculations}
\end{figure}

\subsubsection{Production cross-sections of  $^{142}$Pr}
\label{4.1.14}
We could deduce a few experimental cross-section data for production of the $^{142}$Pr (T$_{1/2}$ = 19.12 h) radioisotope, only obtained directly via (d,2pxn) reactions. The comparison with the theoretical predictions is shown in Fig. 14. 

\begin{figure}
\includegraphics[scale=0.31]{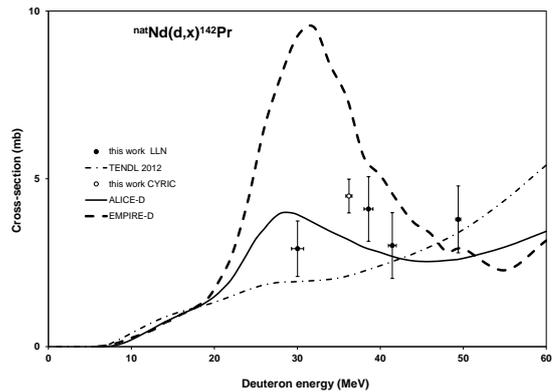}
\caption{Experimental cross-sections for the $^{nat}$Nd(d,x)$^{142}$Pr reaction and comparison with the theoretical code calculations.}
\end{figure}

\subsubsection{Production cross-sections of  $^{139g}$Ce(cum)}
\label{4.1.15}
The direct production cross-sections of the long-lived $^{139}$Ce (T$_{1/2}$ = 137.641 d) via (d,3pxn) reactions are very small. The main contributions to the measured cumulative cross-sections arise from the shorter-lived $^{139}$Pm-$^{139}$Nd-$^{139}$Pr decay chain. Among the used theoretical results the TENDL-2012 data agree rather well with the experimental cumulative excitation function (see Fig. 15).

\begin{figure}
\includegraphics[scale=0.31]{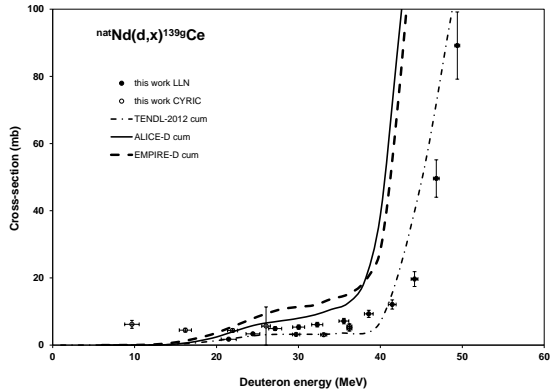}
\caption{Experimental cross-sections for the $^{nat}$Nd(d,x)$^{139g}$Ce reaction and comparison with the theoretical code calculations.}
\end{figure}

\subsection{Integral yields}
\label{4.2}
The integral yields (integrated yield for a given incident energy down to the reaction threshold), calculated from ﬁtted curves to our experimental cross-section data, are shown in Fig. 16-17 in comparison with experimental thick target yields found in the literature. The results are representing so called physical yields (obtained in an instantaneous irradiation time for an incident number of particles having charge equivalent to 1 Coulomb)\cite{19}. Fig. 16 also contains the data from Dmitriev \cite{10} measured at 22 MeV, where the result for $^{144}$Pm is in excellent agreement with our measurement, while his results for $^{143}$Pm and $^{148}$Pm are above and below our curves respectively. 

\begin{figure}
\includegraphics[scale=0.31]{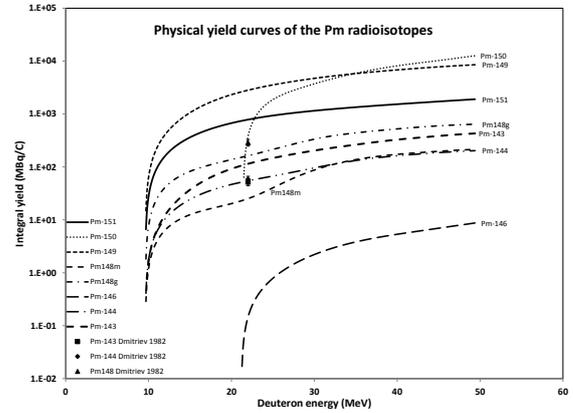}
\caption{Physical yields of the Pm radioisotopes}
\end{figure}

\begin{figure}
\includegraphics[scale=0.31]{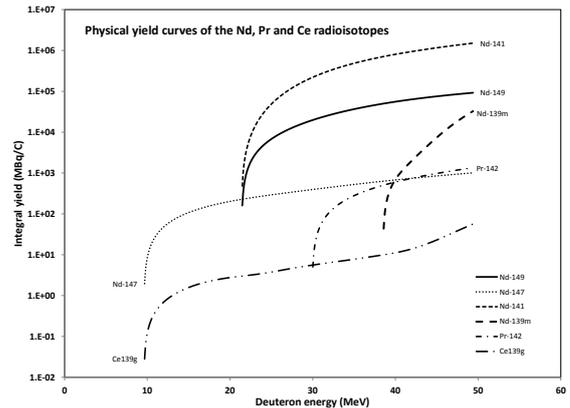}
\caption{Physical yields of the Nd, Pr and Ce radioisotopes}
\end{figure}

\section{Summary and conclusion}
\label{5}
Activation cross-sections of proton induced nuclear reactions on neodymium were measured for the $^{nat}$Nd(d,x) $^{151,150,149,148m,148g,146,144,143}$Pm, $^{149,147,139m}$Nd, $^{142}$Pr and $^{139g}$Ce nuclear reactions. All investigated reactions are presented here for the first time, except the data of \cite{9} for the $^{148}$Nd(d,2n) reaction. Model calculations were done by using the EMPIRE-D and AICE-IPPE-D codes. The data were also compared with the data of the TALYS based TENDL-2012 online library. The predictions of theoretical calculations are only moderately successful. The obtained experimental data provide a basis for upgrading the model codes and input parameters. As it was mentioned the data can be used for selection and optimization of production routes for the medically relevant radioisotopes $^{149}$Pm (T$_{1/2}$ = 53.1 h), $^{141}$Nd (2.49 h) and $^{140}$Nd (T$_{1/2}$=3.37 d).

\begin{table*}[t]
\tiny
\caption{Measured cross-sections for production of the ${}^{151,150,149,148m,148g}$${}^{,}$${}^{146}$${}^{,}$${}^{144}$${}^{,}$${}^{143}$Pm radionuclides}
\centering
\begin{center}
\begin{tabular}{|p{0.1in}|p{0.05in}|p{0.1in}|p{0.1in}|p{0.05in}|p{0.1in}|p{0.1in}|p{0.05in}|p{0.1in}|p{0.1in}|p{0.05in}|p{0.1in}|p{0.1in}|p{0.05in}|p{0.1in}|p{0.1in}|p{0.05in}|p{0.1in}|p{0.1in}|p{0.05in}|p{0.1in}|p{0.1in}|p{0.05in}|p{0.1in}|p{0.1in}|p{0.05in}|p{0.1in}|} \hline 
\multicolumn{3}{|c|}{\textbf{Energy\newline (MeV) }} & \multicolumn{3}{|c|}{${}^{1}$${}^{51}$Pm\textbf{\newline (mb)}} & \multicolumn{3}{|c|}{${}^{1}$${}^{50}$Pm\textbf{\newline (mb)}} & \multicolumn{3}{|c|}{${}^{1}$${}^{49}$Pm\textbf{\newline (mb)}} & \multicolumn{3}{|c|}{${}^{1}$${}^{48m}$Pm\textbf{\newline (mb)}} & \multicolumn{3}{|c|}{${}^{1}$${}^{4}$${}^{8}$${}^{g}$Pm\textbf{\newline (mb)}} & \multicolumn{3}{|c|}{${}^{1}$${}^{46}$Pm\textbf{ \newline (mb)}} & \multicolumn{3}{|c|}{${}^{1}$${}^{44}$Pm\textbf{\newline (mb)}} & \multicolumn{3}{|c|}{${}^{1}$${}^{43}$Pm\textbf{\newline (mb)}} \\ \hline 
\multicolumn{27}{|c|}{\textbf{Sendai, 40 MeV}} \\ \hline 
39.4 & $\pm$ & 0.3 & 4.5 & $\pm$ & 1.1 &  &  &  & 24.4 & $\pm$ & 2.7 & 15.8 & $\pm$ & 1.9 & 5.9 & $\pm$ & 0.8 & 36.2 & $\pm$ & 10.4 & 126 & $\pm$ & 15 & 265 & $\pm$ & 32 \\ \hline 
36.3 & $\pm$ & 0.4 & 3.1 & $\pm$ & 0.4 &  &  &  & 28.2 & $\pm$ & 3.2 & 23.7 & $\pm$ & 2.7 & 6.7 & $\pm$ & 0.8 & 45.7 & $\pm$ & 24.2 & 177 & $\pm$ & 20 & 279 & $\pm$ & 32 \\ \hline 
32.9 & $\pm$ & 0.4 & 2.0 & $\pm$ & 0.5 &  &  &  & 36.3 & $\pm$ & 4.2 & 25.3 & $\pm$ & 3.0 & 6.5 & $\pm$ & 1.0 & 43.9 & $\pm$ & 8.3 & 172 & $\pm$ & 21 & 213 & $\pm$ & 27 \\ \hline 
29.4 & $\pm$ & 0.5 & 2.6 & $\pm$ & 1.0 &  &  &  & 47.5 & $\pm$ & 5.4 & 30.5 & $\pm$ & 3.5 & 8.6 & $\pm$ & 1.7 & 41.4 & $\pm$ & 9.0 & 169 & $\pm$ & 20 & 211 & $\pm$ & 25 \\ \hline 
25.5 & $\pm$ & 0.6 & 2.6 & $\pm$ & 0.6 &  &  &  & 43.7 & $\pm$ & 5.0 & 27.4 & $\pm$ & 3.2 & 8.0 & $\pm$ & 1.1 & 49.7 & $\pm$ & 8.8 & 126 & $\pm$ & 15 & 204 & $\pm$ & 25 \\ \hline 
21.3 & $\pm$ & 0.7 & 4.2 & $\pm$ & 1.0 &  &  &  & 45.4 & $\pm$ & 5.2 & 24.5 & $\pm$ & 2.9 & 9.1 & $\pm$ & 1.2 & 56.9 & $\pm$ & 12.7 & 131 & $\pm$ & 17 & 258 & $\pm$ & 31 \\ \hline 
9.7 & $\pm$ & 1.0 & 14.0 & $\pm$ & 1.6 &  &  &  & 52.7 & $\pm$ & 6.0 & 19.8 & $\pm$ & 2.9 & 16.1 & $\pm$ & 2.0 &  &  &  & 242 & $\pm$ & 28 & 164 & $\pm$ & 20 \\ \hline 
16.2 & $\pm$ & 0.8 & 9.5 & $\pm$ & 1.2 &  &  &  & 72.0 & $\pm$ & 8.2 & 7.4 & $\pm$ & 1.4 & 6.9 & $\pm$ & 1.1 &  &  &  & 205 & $\pm$ & 25 & 348 & $\pm$ & 42 \\ \hline 
\multicolumn{27}{|c|}{\textbf{Louvain la Neuve, 50 MeV}} \\ \hline 
49.4 & $\pm$ & 0.3 & 2.4 & $\pm$ & 0.4 & 3.3 & $\pm$ & 0.4 & 21.4 & $\pm$ & 2.8 & 5.9 & $\pm$ & 0.9 & 2.2 & $\pm$ & 0.5 & 47.2 & $\pm$ & 5.8 & 71 & $\pm$ & 9 & 113 & $\pm$ & 16 \\ \hline 
46.8 & $\pm$ & 0.4 & 3.7 & $\pm$ & 0.5 & 4.3 & $\pm$ & 0.5 & 18.8 & $\pm$ & 2.8 & 9.3 & $\pm$ & 1.2 & 3.0 & $\pm$ & 0.6 & 44.5 & $\pm$ & 5.5 & 74 & $\pm$ & 9 & 160 & $\pm$ & 21 \\ \hline 
44.2 & $\pm$ & 0.4 & 2.7 & $\pm$ & 0.3 & 2.5 & $\pm$ & 0.3 & 27.4 & $\pm$ & 3.6 & 11.8 & $\pm$ & 1.4 & 4.3 & $\pm$ & 0.7 & 45.5 & $\pm$ & 5.6 & 98 & $\pm$ & 12 & 185 & $\pm$ & 23 \\ \hline 
41.4 & $\pm$ & 0.5 & 2.4 & $\pm$ & 0.3 & 2.7 & $\pm$ & 0.3 & 26.1 & $\pm$ & 3.3 & 12.0 & $\pm$ & 1.4 & 5.3 & $\pm$ & 0.8 & 34.7 & $\pm$ & 4.5 & 107 & $\pm$ & 13 & 160 & $\pm$ & 20 \\ \hline 
38.6 & $\pm$ & 0.5 & 2.8 & $\pm$ & 0.4 & 3.5 & $\pm$ & 0.4 & 33.1 & $\pm$ & 4.0 & 15.6 & $\pm$ & 1.8 & 4.9 & $\pm$ & 0.7 & 34.8 & $\pm$ & 4.6 & 116 & $\pm$ & 14 & 199 & $\pm$ & 24 \\ \hline 
35.5 & $\pm$ & 0.6 & 3.5 & $\pm$ & 0.4 & 3.2 & $\pm$ & 0.4 & 33.4 & $\pm$ & 3.9 & 24.4 & $\pm$ & 2.8 & 6.4 & $\pm$ & 0.8 & 47.4 & $\pm$ & 5.9 & 176 & $\pm$ & 20 & 208 & $\pm$ & 24 \\ \hline 
32.3 & $\pm$ & 0.7 & 3.7 & $\pm$ & 0.4 & 3.8 & $\pm$ & 0.4 & 33.0 & $\pm$ & 3.8 & 33.6 & $\pm$ & 3.8 & 9.5 & $\pm$ & 1.1 & 49.1 & $\pm$ & 6.0 & 215 & $\pm$ & 24 & 239 & $\pm$ & 27 \\ \hline 
30.0 & $\pm$ & 0.7 & 3.9 & $\pm$ & 0.4 & 3.7 & $\pm$ & 0.4 & 36.6 & $\pm$ & 4.2 & 40.9 & $\pm$ & 4.6 & 11.1 & $\pm$ & 1.3 & 52.7 & $\pm$ & 6.3 & 214 & $\pm$ & 24 & 260 & $\pm$ & 30 \\ \hline 
27.2 & $\pm$ & 0.8 & 4.5 & $\pm$ & 0.5 & 2.9 & $\pm$ & 0.3 & 39.9 & $\pm$ & 4.6 & 37.9 & $\pm$ & 4.3 & 12.6 & $\pm$ & 1.5 & 50.7 & $\pm$ & 6.1 & 195 & $\pm$ & 22 & 317 & $\pm$ & 36 \\ \hline 
24.4 & $\pm$ & 0.8 & 5.3 & $\pm$ & 0.6 & 3.9 & $\pm$ & 0.4 & 50.9 & $\pm$ & 5.8 & 23.8 & $\pm$ & 2.7 & 10.6 & $\pm$ & 1.2 & 38.6 & $\pm$ & 4.4 & 155 & $\pm$ & 18 & 284 & $\pm$ & 32 \\ \hline 
21.5 & $\pm$ & 0.9 & 6.5 & $\pm$ & 0.7 & 7.6 & $\pm$ & 0.9 & 62.9 & $\pm$ & 7.1 & 8.8 & $\pm$ & 1.0 & 5.9 & $\pm$ & 0.7 & 31.6 & $\pm$ & 3.8 & 144 & $\pm$ & 16 & 321 & $\pm$ & 36 \\ \hline 
\end{tabular}
\end{center}
\end{table*}

\begin{table*}[t]
\tiny
\caption{Measured cross-sections for production of the ${}^{149}$${}^{,}$${}^{147,139m}$Nd, ${}^{142}$Pr and ${}^{139g}$Ce radionuclides}
\centering
\begin{center}
\begin{tabular}{|p{0.1in}|p{0.05in}|p{0.1in}|p{0.2in}|p{0.1in}|p{0.2in}|p{0.2in}|p{0.1in}|p{0.2in}|p{0.2in}|p{0.1in}|p{0.2in}|p{0.2in}|p{0.1in}|p{0.2in}|p{0.2in}|p{0.1in}|p{0.2in}|p{0.2in}|p{0.1in}|p{0.2in}|} \hline 
\multicolumn{3}{|c|}{\textbf{Energy\newline (MeV) }} & \multicolumn{3}{|c|}{${}^{1}$${}^{49}$Nd\textbf{\newline (mb)}} & \multicolumn{3}{|c|}{${}^{1}$${}^{47}$Nd\textbf{\newline (mb)}} & \multicolumn{3}{|c|}{${}^{1}$${}^{41}$Nd\textbf{\newline  (mb)}} & \multicolumn{3}{|c|}{${}^{1}$${}^{39m}$Nd\textbf{\newline  (mb)}} & \multicolumn{3}{|c|}{${}^{1}$${}^{42}$Pr\textbf{\newline  (mb)}} & \multicolumn{3}{|c|}{${}^{1}$${}^{39g}$Ce\textbf{\newline (mb)}} \\ \hline 
\multicolumn{21}{|c|}{\textbf{Sendai, 40 MeV}} \\ \hline 
39.4 & $\pm$ & 0.3 &  &  &  & 20.5 & $\pm$ & 2.6 &  &  &  &  &  &  &  &  &  & 5.6 & $\pm$ & 0.8 \\ \hline 
36.3 & $\pm$ & 0.4 &  &  &  & 23.2 & $\pm$ & 2.6 &  &  &  &  &  &  &  &  &  & 4.9 & $\pm$ & 0.8 \\ \hline 
32.9 & $\pm$ & 0.4 &  &  &  & 18.4 & $\pm$ & 2.4 &  &  &  &  &  &  &  &  &  & 3.1 & $\pm$ & 0.5 \\ \hline 
29.4 & $\pm$ & 0.5 &  &  &  & 16.3 & $\pm$ & 2.1 &  &  &  &  &  &  &  &  &  & 3.2 & $\pm$ & 0.4 \\ \hline 
25.5 & $\pm$ & 0.6 &  &  &  & 14.0 & $\pm$ & 1.8 &  &  &  &  &  &  &  &  &  & 5.7 & $\pm$ & 5.7 \\ \hline 
21.3 & $\pm$ & 0.7 &  &  &  & 15.8 & $\pm$ & 2.1 &  &  &  &  &  &  &  &  &  & 4.3 & $\pm$ & 0.5 \\ \hline 
9.7 & $\pm$ & 1.0 &  &  &  & 39.0 & $\pm$ & 4.6 &  &  &  &  &  &  &  &  &  & 6.2 & $\pm$ & 1.2 \\ \hline 
16.2 & $\pm$ & 0.8 &  &  &  & 26.1 & $\pm$ & 3.4 &  &  &  &  &  &  &  &  &  & 4.4 & $\pm$ & 0.6 \\ \hline 
\multicolumn{21}{|c|}{\textbf{Louvain la Neuve, 50 MeV}} \\ \hline 
49.4 & $\pm$ & 0.3 & 15.0 & $\pm$ & 1.7 & 22.5 & $\pm$ & 2.5 & 478 & $\pm$ & 55 & 11.82 & $\pm$ & 1.35 & 3.79 & $\pm$ & 1.00 & 89.2 & $\pm$ & 10.0 \\ \hline 
46.8 & $\pm$ & 0.4 & 15.1 & $\pm$ & 1.7 & 22.1 & $\pm$ & 2.5 & 452 & $\pm$ & 52 & 5.76 & $\pm$ & 0.65 & ~ &  &  & 49.6 & $\pm$ & 5.6 \\ \hline 
44.2 & $\pm$ & 0.4 & 15.5 & $\pm$ & 1.7 & 22.2 & $\pm$ & 2.5 & 429 & $\pm$ & 49 & 2.50 & $\pm$ & 0.31 & ~ &  &  & 19.7 & $\pm$ & 2.2 \\ \hline 
41.4 & $\pm$ & 0.5 & 17.4 & $\pm$ & 2.0 & 22.7 & $\pm$ & 2.6 & 411 & $\pm$ & 47 & 0.89 & $\pm$ & 0.14 & 3.01 & $\pm$ & 0.98 & 12.1 & $\pm$ & 1.4 \\ \hline 
38.6 & $\pm$ & 0.5 & 18.5 & $\pm$ & 2.1 & 22.4 & $\pm$ & 2.5 & 427 & $\pm$ & 49 & 0.59 & $\pm$ & 0.10 & 4.10 & $\pm$ & 0.96 & 9.3 & $\pm$ & 1.1 \\ \hline 
35.5 & $\pm$ & 0.6 & 18.7 & $\pm$ & 2.1 & 23.4 & $\pm$ & 2.6 & 426 & $\pm$ & 49 &  &  &  & ~ &  &  & 7.1 & $\pm$ & 0.8 \\ \hline 
32.3 & $\pm$ & 0.7 & 17.9 & $\pm$ & 2.0 & 23.2 & $\pm$ & 2.6 & 406 & $\pm$ & 47 &  &  &  & ~ &  &  & 6.1 & $\pm$ & 0.7 \\ \hline 
30.0 & $\pm$ & 0.7 & 17.4 & $\pm$ & 2.0 & 23.1 & $\pm$ & 2.6 & 393 & $\pm$ & 45 &  &  &  & 2.92 & $\pm$ & 0.83 & 5.4 & $\pm$ & 0.6 \\ \hline 
27.2 & $\pm$ & 0.8 & 15.4 & $\pm$ & 1.7 & 21.7 & $\pm$ & 2.4 & 298 & $\pm$ & 35 &  &  &  &  &  &  & 4.9 & $\pm$ & 0.6 \\ \hline 
24.4 & $\pm$ & 0.8 & 12.4 & $\pm$ & 1.4 & 20.8 & $\pm$ & 2.3 & 197 & $\pm$ & 23 &  &  &  &  &  &  & 3.3 & $\pm$ & 0.4 \\ \hline 
21.5 & $\pm$ & 0.9 & 11.1 & $\pm$ & 1.2 & 20.5 & $\pm$ & 2.3 & 46 & $\pm$ & 6 &  &  &  &  &  &  & 1.7 & $\pm$ & 0.2 \\ \hline 
\end{tabular}
\end{center}
\end{table*}

\section{Acknowledgements}
\label{6}

This work was done in the frame MTA-FWO research project and ATOMKI-CYRIC collaboration. The authors acknowledge the support of research projects and of their respective institutions in providing the materials and the facilities for this work. 
 



\bibliographystyle{elsarticle-num}
\bibliography{Ndd}







\end{document}